\documentclass[10pt, conference, letterpaper]{ IEEEtran}

\IEEEoverridecommandlockouts

\usepackage{subcaption} 
\usepackage{caption}
\usepackage{float}
\usepackage{enumitem}
\usepackage{balance}
\usepackage{cite}
\usepackage{amsmath,amssymb,amsfonts}
\usepackage{graphicx}
\usepackage{xcolor}
\usepackage{stfloats}
\usepackage{amsthm}
\usepackage{amsmath}
\DeclareMathOperator*{\argmin}{arg\,min}
\makeatletter%
\if@twocolumn%
\newcommand{\Figwidth}{\columnwidth}%
\else
\newcommand{\Figwidth}{4.5in}%
\fi%
\makeatother%

\allowdisplaybreaks

\renewcommand{\Figwidth}{3.5in}

\def\log{{\rm{log}}}

\def\bh{\mathbf{h}}

\def\bw{\mathbf{w}}

\def\by{\mathbf{y}}

\def\bX{\mathbf{X}}
\def\bx{\mathbf{x}}

\def\bI{\mathbf{I}}
\def\be{\mathbf{e}}

\def\bA{\mathbf{A}}

\def\bf{\mathbf{f}}

\def\bA{\mathbf{A}}

\def\btheta{\boldsymbol{\theta}}

\def\bbeta{\boldsymbol{\beta}}

\newtheorem{remark}{Remark}
\newtheorem{theorem}{Theorem}
\renewcommand{\proof}{{\it{Proof:}}}

\usepackage{balance}

\begin{document}

\title{Coherence-Aware Distributed Learning under Heterogeneous Downlink Impairments}

\author{\IEEEauthorblockN{Mehdi Karbalayghareh, David J. Love, and Christopher G. Brinton}
}

\maketitle

\begin{abstract}
The performance of federated learning (FL) over wireless networks critically depends on accurate and timely channel state information (CSI) across distributed devices. This requirement is tightly linked to how rapidly the channel gains vary, i.e., the coherence intervals. In practice, edge devices often exhibit unequal coherence times due to differences in mobility and scattering environments, leading to unequal demands for pilot signaling and channel estimation resources. Conventional FL schemes that overlook this {\em coherence disparity} can suffer from severe communication inefficiencies and training overhead. This paper proposes a coherence-aware, communication-efficient framework for joint channel training and model updating in practical wireless FL systems operating under heterogeneous fading dynamics. Focusing on downlink impairments, we introduce a resource-reuse strategy based on {\em product superposition}, enabling the parameter server to efficiently schedule both static and dynamic devices by embedding global model updates for static devices within pilot transmissions intended for mobile devices. We theoretically analyze the convergence behavior of the proposed scheme and quantify its gains in expected communication efficiency and training accuracy. Experiments demonstrate the effectiveness of the proposed framework under mobility-induced dynamics and offer useful insights for the practical deployment of FL over wireless channels.

\end{abstract}


\IEEEpeerreviewmaketitle
\section{Introduction}
\label{sec:introduction}
Federated learning (FL) facilitates decentralized model training across edge devices without the need to exchange raw data~\cite{pmlr-v54-mcmahan17a,Li_FedSurvey_SPM2020,Nguyen_JSAC2021}. Although FL is promising in terms of privacy and scalability, it often encounters significant communication constraints over wireless networks~\cite{Ahmed_INFOCOM25_ComEfficient}. Its efficacy hinges on the availability of a reliable communication system and the ability to accurately acquire and exchange link qualities (channel state information, CSI) among nodes. The quality of this CSI is in turn tightly linked to how rapidly the channel gains vary, quantified by the \textit{coherence interval}.

While most existing FL frameworks assume uniform fading rates (i.e., equal coherence intervals) across all devices, real-world wireless networks rarely conform to this assumption. Variations in node mobility and scattering environments lead to the {\em coherence disparity}, where devices experience unequal coherence times (e.g., the coexistence of low-mobility and high-mobility devices~\cite{Lin_JSAC_HighLowMobility}). This disparity degrades both downlink model delivery and uplink gradient aggregation quality, rendering conventional communication strategies inefficient.

In the downlink channel estimation phase, a common pilot signal is shared among all receivers, resulting in uniform time and power allocation across devices~\cite{Tong2004SigProcMag,Lozano2010EUWConf}. This is true even when the links have different coherence times. When some channels vary more rapidly than others, the pilot sequence that is geared toward some links may be either inadequate or excessive for other links. Enforcing strict orthogonality between pilot and data transmission further amplifies this inefficiency, as it increases overhead and reduces the time available for sending model updates. In fact, under severe coherence disparity, even static devices—which have less trouble participating in FL rounds—may fail to receive the full model due to bandwidth wasted on redundant pilots. This can significantly increase bias in the learned parameters and degrade overall FL performance.

To address this challenge, {\em pilot reuse} must be integrated with the FL framework to ensure both bandwidth and resource efficiency, as well as efficient device scheduling and model delivery. Recent work has shown that the most effective pilot reuse technique under coherence disparity is {\em product superposition}~\cite{Mehdi_TWC24_CoherenceDis,Fadel2016CoherenceDisparity}, which overlays data for slow fading users onto pilot symbols intended for fast fading users (i.e., overlapping pilot and data transmission), enabling simultaneous pilot and data transmission within the same timeslots. Originally developed for Multiple Input Multiple Output (MIMO) downlink systems, this method allows fast users to obtain fresh pilots as often as needed while slow users exploit unused pilot capacity to receive model parameters at minimal additional cost. Coupling this strategy with coherence-aware device scheduling has the potential to significantly reduce overhead while guaranteeing timely delivery of full model updates, and ensuring that all devices—including static ones—can remain active participants in the FL process.


\subsection{Related Works}
Communication efficiency and reliability have been widely acknowledged as critical bottlenecks in practical wireless FL, as high-dimensional model updates must be exchanged over bandwidth-constrained and noisy wireless channels~\cite{Lim_ComSurvey2020,Hu_ComSurvey2021,Amiri_TWC2021_FLConvergence,Ahn2020FedDistill}. In the FL literature, the majority of studies have primarily focused on {\em uplink communication} imperfection and tried to reduce the overhead from devices to the PS, proposing two main techniques. The first is digital FL, which allocates orthogonal resource blocks to each device so that the PS can decode and aggregate local gradients individually. The second is over-the-air (OTA) FL, which utilizes the superposition property of the wireless multiple-access channel to perform simultaneous analog transmissions, enabling one-shot gradient aggregation. While digital FL emphasizes efficient scheduling and bandwidth allocation~\cite{Shi2021JointScheduling,Yang2020Scheduling,Wang2022QuantizedFL,Bouzinis2023WirelessQFL,Salehi2021FLUnreliable}, OTA FL focuses on power control mechanisms to mitigate aggregation noise~\cite{Yang2020OTAComp,Amiri2020FLFading,Zhu2021OneBitOTA,Zhu2020Broadband,Michelusi2024NonCoherent,Sery2021OTAFLHeterogeneous,Wang_INFOCOM25_delayedCSI}. However, both lines of work typically assume relatively homogeneous wireless conditions—not only in terms of path loss, but also in terms of channel coherence conditions, where all devices are presumed to experience similar fading dynamics. 

There are relatively fewer studies focused on imperfect downlink transmission, i.e., for broadcasting the global FL model and its impact on system performance. Amiri~{\em et al.}~\cite{Amiri2021NoisyDL} studied the performance of FL over noisy downlink channels, proposing analog (unquantized) and digital (quantized) model broadcasting from the parameter server (PS) to devices with imperfect CSI used for decoding. Building on this direction, Park~{\em et al.}~\cite{Park2021NoisyFeedback} considered feedback imperfections in the form of noisy and limited-rate links during downlink transmission, analyzing their effect on the convergence of distributed gradient methods. Similarly, Nguyen~{\em et al.}~\cite{Nguyen2021CSIUncertainty} studied the impact of uncertain CSI on downlink transmission, proposing robust aggregation schemes for FL in the presence of imperfect CSI at the devices. Cui~{\em et al.}~\cite{Cui2022Beamforming} addressed downlink imperfections by proposing a joint beamforming strategy at the PS to improve model aggregation quality under fading. Addressing communication efficiency, Caldas~{\em et al.}~\cite{Caldas2018ReducingClients} introduced a system that reduces the downlink communication load by selectively distributing compressed model updates tailored to client resource constraints. Along similar lines, Tang~{\em et al.}~\cite{Tang2019DoubleSqueeze} proposed an error-compensated compression scheme where the downlink model is doubly compressed using stochastic gradient and memory error correction techniques, mitigating the impact of bandwidth constraints on FL performance.

While all the aforementioned works consider downlink transmission imperfections in FL networks, they overlook another critical factor: {\em coherence disparity,} where devices experience unequal channel coherence times due to mobility and environmental heterogeneity. Such mismatches result in uneven pilot requirements and bandwidth inefficiencies, which can degrade global model delivery, especially for fast-fading devices, and waste resources of static devices. 

\subsection{Contributions}
Motivated by this, we study FL systems under downlink coherence disparity, proposing a product superposition-based downlink model transmission and device scheduling framework. We analyze the effectiveness of our approach in enhancing communication efficiency and reliability through careful design of overlapping pilot and parameter transmission in the downlink. As the first work to address FL under coherence disparity, we focus solely on the downlink, and leave the uplink analysis for an extension. Our proposed scheme and theoretical results offer valuable insights for the practical implementation of FL over wireless networks, where heterogeneous coherence conditions across links are pervasive.

The main contributions of this paper are as follows:
\begin{itemize}[leftmargin=5mm]
    \item We introduce a coherence-aware FL system model that captures downlink heterogeneity due to the coexistence of static and dynamic devices with unequal coherence times. Our model harmonizes pilot reuse techniques with FL system design, paving the way for more bandwidth-efficient learning under coherence disparity.
    \item We employ product superposition to enable overlapping pilot and parameter transmission in the downlink. This allows static devices to reuse pilot slots for receiving the global model, while dynamic devices can coherently decode the partial model by estimating their respective {\em virtual channels}, which is the product of their own link gain and the parameter signal intended for static devices.
    \item We propose coherence-aware device scheduling and adaptive gradient aggregation strategies to address partial model reception. We explore two aggregation methods for dynamic devices: Zero-Filling (ZF), which substitutes missing parameters with zeros, and Previous Local Model Filling (PLMF), which reuses prior local model entries.
    \item We provide a convergence analysis of the proposed scheme under imperfect CSI, capturing the impact of estimation errors and fading mismatch on learning performance.
\end{itemize}

\section{System Model}
\label{sec:system-model}
\begin{figure}
    \centering
    \includegraphics[page=1, width=\Figwidth]{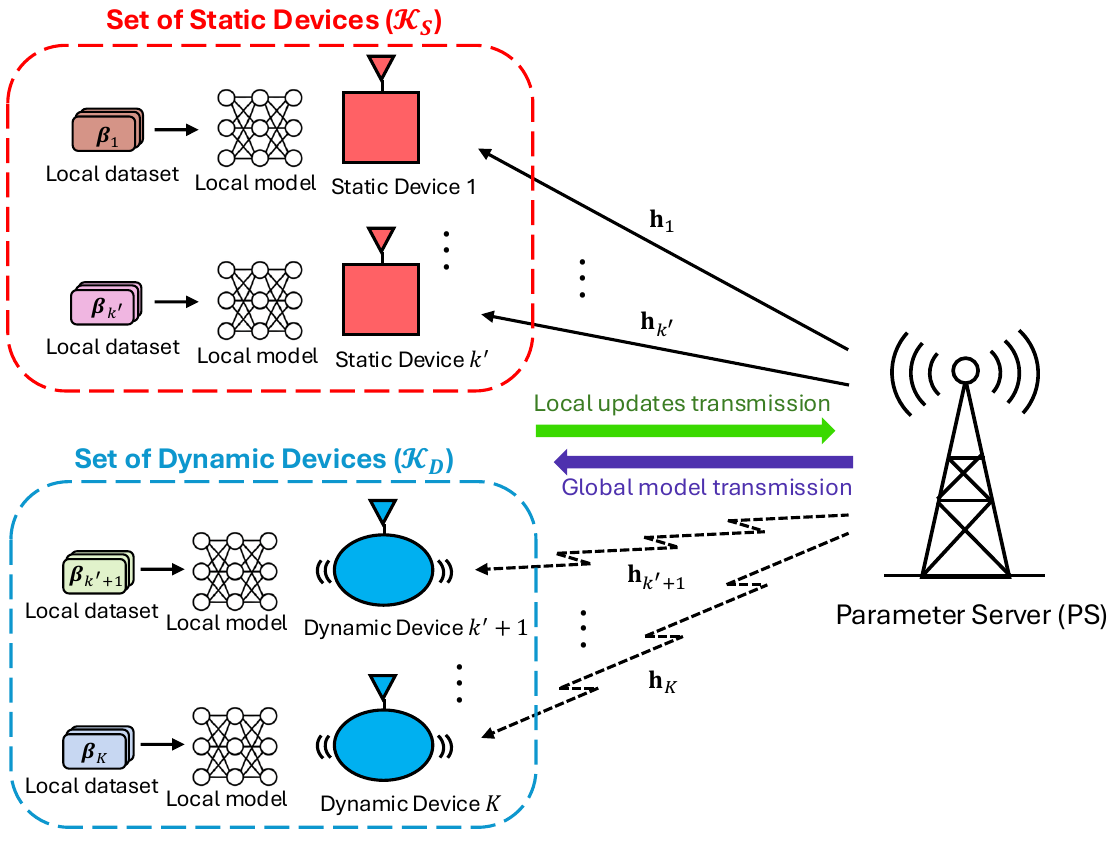}
    \caption{Wireless FL scenario considered. Our focus is on studying and mitigating the impact of downlink impairments. \vspace{-0.2in}}
    \label{fig:scenario}
\end{figure}

We consider an FL system with a PS and $K$ edge devices, depicted in Fig.~\ref{fig:scenario}. Throughout the paper, we will use the terms ``device'', ``receiver'', and ``user'' interchangeably to denote the edge terminals in the downlink. Each device $k \in [K] = \{1, \ldots, K\}$ possesses a local dataset $\mathcal{B}_k$ with cardinality $B_k = |\mathcal{B}_k|$ datapoints. Let $B \triangleq \sum_{k=1}^{K} B_k$, and $F_k(\boldsymbol{\theta}) \triangleq \frac{1}{B_k} \sum_{\boldsymbol{v} \in \mathcal{B}_k} f(\boldsymbol{\theta}, \boldsymbol{v})$ denote the local loss at device~$k$, where $f$ is the empirical loss function. For a $d$-dimensional global model denoted by $\btheta \in \mathbb{R}^d$, the global loss function to be minimized is
\begin{align}
\label{eq:global-loss}
   F(\btheta) = \sum_{k=1}^{K} \frac{B_k}{B} F_k(\btheta).
\end{align}

FL aims to minimize $F(\btheta)$ through iterative collaboration between the PS and $K$ edge devices. In each iteration~$t$, the PS broadcasts the global model $\btheta^{(t)}$ to the devices over wireless channels, which are subject to impairments such as fading, noise, and decoding errors. Consequently, each device~$k$ receives an imperfect version of the model, denoted by ${\bar{\btheta}}_k^{(t)}$. Each device then performs $\tau$ steps of stochastic gradient descent (SGD) using its local data. At step~$i$, it selects a random minibatch $\bbeta_{k,i}^{(t)}$ and updates its model via
\begin{align}
\btheta_{k,i+1}^{(t)} = \btheta_{k,i}^{(t)} - \eta_{k,i}^{(t)} \nabla F_k \Big(\btheta_{k,i}^{(t)}, \bbeta_{k,i}^{(t)}\Big), \quad i \in [\tau],
\label{eq:local-updates}
\end{align}
where $\btheta_{k,1}^{(t)} = \bar{\btheta}_k^{(t)}$ and $\eta_{k,i}^{(t)}$ is the learning rate. After local updates, device~$k$ sends $\Delta \btheta_k^{(t)} = \btheta_{k,\tau}^{(t)} - \btheta_{k,1}^{(t)}$ to the PS, which receives a noisy estimate $\widehat{\Delta{\btheta}}_k^{(t)}$. The PS aggregates these updates to refine the global model as
\begin{align}
\btheta^{(t+1)} = \btheta^{(t)} + \sum_{k=1}^{K} \frac{B_k}{B} \widehat{\Delta{\btheta}}_k^{(t)}.
\label{eq:model-update}
\end{align}
This process continues until convergence over $t = 1,...,T$ training rounds.

\subsection{Channel Model}
We assume that the PS is equipped with $M$ antennas, while each device has a single antenna. The channel vector from the PS to device $k$ is denoted by $\bh_k \in \mathbb{C}^{M \times 1}$, which has independent identically distributed
(i.i.d.) entries $\mathcal{CN} (0,1)$. The system operates under frequency-flat channels and a block-fading model, where $\bh_k$ remains constant over $T_k$ symbols and changes independently across blocks. Due to differences in mobility (i.e., fading dynamics), the coherence times $T_k$ are not identical across devices. As illustrated in Fig.~\ref{fig:scenario}, we partition the devices into a set of static devices, denoted by $\mathcal{K}_S$, and a set of dynamic devices, denoted by $\mathcal{K}_D$. For any $k, k'' \in \mathcal{K}_D$ with $k \neq k''$, it holds that $T_k \neq T_{k''}$. This coherence disparity is addressed in detail in Section~\ref{sec:communication-protocol}, where we propose an efficient strategy for the joint participation of both static and dynamic devices in the FL system.

Let $\bX(t) \triangleq [\bx_1(t), \ldots, \bx_M(t)]^T$ denote the signal transmitted by the PS in iteration~$t$ across its $M$ antennas, where each $\bx_i(t) \in \mathbb{C}^{T_c \times 1}$ represents the signal vector transmitted by antenna~$i$ over $T_c$ time slots (symbols). Then, the received signal at device~$k$ is
\begin{align}
    \by_k(t)=\bh_k^H(t) \,\bX(t) + \bw_k(t),\,\, k=1, \cdots, K,
    \label{eq:received-signal}
\end{align}
where $\bw_k(t) \in \mathbb{C}^{T_c}$ denotes the additive Gaussian noise vector at device~$k$ whose elements are i.i.d. with zero mean and variance $\sigma_w^2$. 
The PS is assumed to have an average power constraint $\rho$, i.e.,
\begin{align}
    \mathbb{E}\Big[\sum_{i=1}^{M}{\rm tr}\big(\bx_i \bx^H_i\big)\Big] \leq \rho T_c.
    \label{eq:PSpower-constraint}
\end{align}

\begin{remark}
For dynamic devices, the instantaneous downlink channels $\bh_k$ are unknown to both the devices and the PS, whereas the channels of static devices are assumed to be perfectly known at both ends due to their long coherence times. We focus on the frequency-division duplexing (FDD) framework, which facilitates clearest presentation of the proposed ideas and computations. In this setting, one frequency band is allocated for the downlink—carrying product-superposed pilots and global model parameters—while a separate band is used for the uplink, transmitting the aggregated gradients. The proposed framework and accompanying analyses extend to time-division duplexing (TDD) systems with minor modifications in pilot placement, which are omitted here for brevity.
\end{remark}

\subsection{Baseline Transmission Scheme}
The baseline scheme employs orthogonal pilot and data (model parameters) transmissions in the downlink, without using superposition pilots—a method shown to be highly inefficient under {\em coherence disparity}~\cite{Mehdi_TWC24_CoherenceDis}. This conventional approach fails to account for the unequal link requirements across devices, which are typical in practical FL over wireless channels. As a result, dynamic devices experience either degraded channel estimation accuracy or reduced time for data reception (leading to incomplete model updates), both of which negatively impact FL performance. Furthermore, this inefficient use of wireless resources results in significant communication overhead. During each iteration, the downlink transmission block consists of a pilot phase followed by a data phase. The PS first transmits an orthogonal pilot matrix for channel estimation and then uses the remaining portion of the block to transmit the model parameters. The transmit signal is given by
\begin{align}
    \bX(t) = \big[\sqrt{\rho_p}\, \bX_p\,, \, \sqrt{\rho_d}\, \bX_d(t) \big],
\end{align}
where $\bX_p \in \mathbb{C}^{M \times M}$ is a unitary pilot matrix such that $\bX_p \bX_p^H = \bI$ and is independent of~$t$.  $\bX_d(t) \in \mathbb{C}^{M \times (T_c-M)}$ is the data matrix, sent over $T_c-M$ data slots within the length-$T_c$ coherence interval. $\rho_p$ and $\rho_d$ are the average power used for channel training and data, respectively, and satisfy the power constraint in \eqref{eq:PSpower-constraint} as
\begin{align}
\rho_p M + \rho_d (T_c - M) \leq \rho T_c. 
\label{eq:power-constarint-conventional}
\end{align}

\section{Coherence-Aware Device Scheduling and Communication Protocol}
\label{sec:communication-protocol}
In this section, we detail our coherence-aware device scheduling and proposed communication protocol. This yields efficient implementation of wireless FL under unequal coherence intervals, where channel estimation requirements are not uniform across devices.
As a result, not all devices, or a randomly selected subset, can participate in each FL iteration. 

In each communication round $t$, the PS selects $K$ devices to participate in distributed learning. Let $\mathcal{K}_S \triangleq \{1, \ldots, k'\}$, with $|\mathcal{K}_S| = K' \leq K$, denote the set of static devices (with consistently stable channels and access to accurate CSI), who are always eligible to participate. To complete the set of $K$ participating devices, the PS then selects $K - K'$ dynamic devices, whose channels may have changed since the last estimate\footnote{The PS has the knowledge of coherence times at the time of scheduling. This information can be shared over the uplink with negligible overhead, and we omit further discussion for brevity.}. We denote the set of dynamic devices participating in the training by $\mathcal{K}_D \triangleq \{k'+1, \ldots, K\}$.

Due to the coexistence of dynamic and static devices, downlink transmission in each iteration must be carefully designed to serve a dual purpose: enabling channel estimation and coherent partial model delivery for dynamic devices, while simultaneously delivering the global model to static devices.

\subsection{Downlink Signaling: Integrated Pilot-Parameter Broadcast}
\label{sec:downlink-signaling}

Without loss of generality, we order the dynamic coherence intervals in descending order: $T_{k'+1} >\ldots > T_K$. This implies that device~$K$ experiences the fastest fading speed, and consequently, its coherence time determines the pilot duty cycle in the downlink signaling design. Assume that $s$ symbols are required to share the full global model in the downlink\footnote{The value of $s$ depends on the transmission mode (analog/digital), modulation scheme, coding rate, and quantization level, which together determine its relationship with model dimension $d$. A detailed treatment of these factors is beyond the scope of this work; some of them are discussed in~\cite{Amiri2020FLFading}.}. For simplicity of exposition and analytical tractability, we assume that $s=q T_K, q \in Z$, and that the PS begins transmitting the parameters at the start of device~$K$'s coherence interval (coherence interval information is known at the PS). These assumptions can be relaxed within our proposed scheme (see Remark~\ref{re:coherence-misalignment}). In iteration $t$, all $s$ symbols are transmitted over $q$ sub-blocks of length~$T_K$. The transmitted super-symbols in sub-block~$q' \in \{1, \dots, q\}$ is 
\begin{align}
    \bX_{q'}(t) = \Big[\sqrt{\rho_p}\, \bX_{p,q'}^\theta(t) \, \bX_p \,, \, \sqrt{\rho_d}\, \bX_{p,q'}^\theta(t)\,  \bX_{d,q'}^\theta(t) \Big],
    \label{eq:ps-transmit}
\end{align}
where $\mathbf{X}_p \in \mathbb{C}^{M \times M}$ is a unitary pilot matrix that remains fixed across all sub-blocks. $\bX_{p,q'}^\theta(t) \in \mathbb{C}^{M \times M}$ denotes a partial parameter matrix containing the first $M$ model symbols in sub-block~$q'$, transmitted via the $M$ antennas during the pilot phase of the interval. $\bX_{d,q'}^\theta(t) \in \mathbb{C}^{M \times (T_K-M)}$ denotes the partial parameter matrix containing the remaining $T_K-M$ model symbols in sub-block~$q'$, transmitted via the $M$ antennas during the data phase of the interval.

By substituting \eqref{eq:ps-transmit} into \eqref{eq:received-signal}, the received signal at all devices can be obtained. Any static device $k \in \mathcal{K}_S$ can directly decode both $\bX_{p,q'}^\theta(t)$ and $\bX_{d,q'}^\theta(t)$ during the pilot and data phases of sub-block~$q'$, respectively, since it knows both $\bh_k$ and $\bX_p$. The same holds for any dynamic device whose channel has remained unchanged since its last estimate. However, device~$K$ (fastest link), as well as any other dynamic device whose channel has changed, must first estimate its equivalent channel $\bh_k^H \bX_{p,q'}^\theta(t)$, i.e., the product of its link gain with the partial parameter matrix, during the pilot phase. It then uses this estimate to coherently decode $\bX_{d,q'}^\theta(t) $ during the data phase. This strategy enables full model delivery to static devices and efficient partial model delivery to dynamic devices, while reducing overall communication overhead. Let $\bf_k \triangleq \bh_{k}^H \bX_{p,q'}^\theta$. Then, the MMSE estimate of $\bf_k$ is denoted $\overline{\bf}_k$~\cite{Hassibi2003HowMuch}:
\begin{align}
\label{eq:MMSE-equivalent}
   \overline{\bf}_k =&\, \mathbb{E}\big[\bf_k\,\by_k^H\big]\mathbb{E}\big[\by_k\,\by_k^H\big]^{-1} \by_k \nonumber\\
   =& \,{\frac{M \rho_p}{M \rho_p+\sigma_w^2}}\big(\bf_k+\bw_k\big).   
\end{align}
The estimation error is denoted $\tilde{\bf}_k = \bf_k - \overline{\bf}_k$, which is Gaussian with covariance $\sigma_{e,k}^2\bI$, where
\begin{align}
    \sigma_{e,k}^2 = \frac{M\sigma_w^2}{M \rho_p+\sigma_w^2}.
    \label{eq:estimation-error}
\end{align}

\begin{figure}
    \centering
    \includegraphics[page=2, width=\Figwidth]{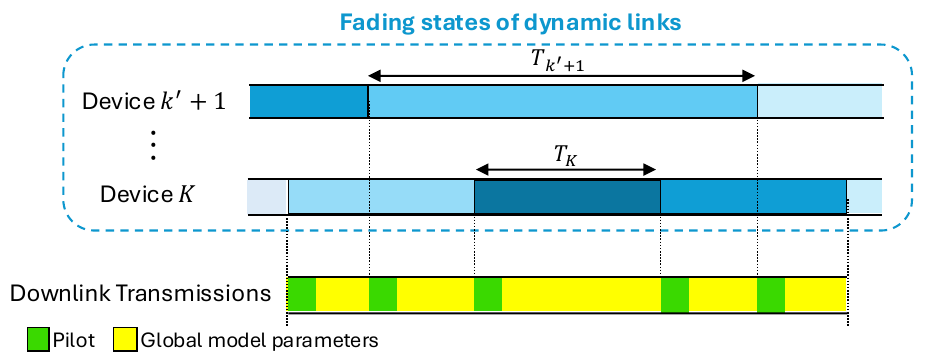}
    \caption{Heterogeneous link coherence intervals. \vspace{-0.15in}}
    \label{fig:coherence-intervals}
\end{figure}

\begin{remark}
    \label{re:coherence-misalignment}
    While we assume that $q$ is an integer and the downlink signaling is aligned with the start of device~$K$'s coherence interval, the proposed scheme can accommodate non-integer values of $q$ and signaling misaligned with the shortest coherence interval. All signaling design takes place at the PS, where both the coherence times and the value of $s$ are known. The PS can apply the same transmission principles described in Eq.~\eqref{eq:ps-transmit} for any block size. Moreover, it can flexibly shift the superimposed pilots within a block to align with other devices' coherence intervals. It is also possible to insert multiple product superpositions within a single block, as it results in non-interfering pilots and model parameters. An example of this scenario is illustrated in Fig.~\ref{fig:coherence-intervals}, where the green regions indicate valid pilot positions for applying product superposition.
\end{remark}

\subsection{Pilot-Parameter Power Allocation}
The proposed signaling in Eq.~\eqref{eq:ps-transmit} must satisfy the total power constraint $\rho$ at the PS, as defined in Eqs.~\eqref{eq:PSpower-constraint} and \eqref{eq:power-constarint-conventional}
\begin{align}
    qM(\rho_p + \rho_d(T_K - M)) \leq \rho s.
\end{align}
Over each sub-block~$q'$, our scheme carries data (model parameters) to a static device~$k' \in \mathcal{K}_S$ during the pilot phase (first $M$ time slots), resulting in the additional rate
\begin{align}
    R_{k', q'} = \frac{M}{T_K} \mathbb{E}\left[\log_2 \left(1 + \frac{\rho_p}{M\sigma_w^2} \bh_{k'}^H \bh_{k'}\right)\right]. \label{eq:rate_static}
\end{align}
However, a dynamic device~$k \in \mathcal{K}_D$, whose channel has changed, can only receive data during the data phase (remaining $T_K-M$ time slots). Therefore, the achievable rate is
\begin{align}
    R_{k, q'} = \left(1 - \frac{M}{T_K}\right) \mathbb{E}\left[\log_2\left(1 + \gamma_{k,q'} \,\overline{\bf}_{k, q'}^H \overline{\bf}_{k, q'}\right)\right] ,\label{eq:rate_dynamic}
\end{align}
where $\overline{\bf}_{k, q'}$ denotes the estimate of the equivalent channel for dynamic device~$k$ at sub-block~$q'$ (see Eq.~\eqref{eq:MMSE-equivalent} for details), and $\gamma_{k,q'}$ denotes its effective signal-to-noise ratio (SNR). 

Since the proposed scheme reuses the pilot slots of {\em dynamic devices} to deliver data to static devices, we focus on a power allocation that maximizes the dynamic devices’ rate, ensuring reliable communication even though they receive only partial model parameters\footnote{Dynamic links are the bottleneck in the system.}. Maximizing the achievable rate at dynamic devices in Eq.~\eqref{eq:rate_dynamic} yields the following optimal power allocation:
\begin{align}
    \rho_d^* &= \frac{\sigma_w^2+\rho T_K}{M\sqrt{T_K-M}(1 + \sqrt{T_K-M})} \label{eq:rho_d_opt} \\
    \rho_p^* &= \frac{\rho T_K}{M} - \rho_d^*(T_K - M) \label{eq:rho_p_opt}
\end{align}

These derivations are given in Appendix~\ref{app:proof-PowerAllocation}.

\subsection{Uplink Model Aggregation}
The model update at the PS after each communication round is given in Eq.~\eqref{eq:model-update}. In this work, we adopt a simplified uplink model under FDD mode (devices transmit their local updates over orthogonal channels) to focus on the effects of downlink imperfections stemming from {\em coherence disparity}. In particular, we aim to design communication-efficient signaling schemes that enable overlapping pilot and parameter transmission under heterogeneous coherence intervals, which introduces new challenges such as imperfect CSI estimation and partial model delivery. By addressing these challenges, our goal is to support efficient and robust federated learning in practical environments with both static and dynamic devices. To isolate the downlink impairments, we assume perfect CSI in the uplink and reliable transmission of local updates. This abstraction allows us to analyze the core issues introduced by downlink-side limitations and evaluate their impact on training performance and convergence. While our framework remains compatible with different aggregation strategies under imperfect CSI, a comprehensive treatment of uplink-side challenges under unequal fading dynamics is deferred to the extended journal version due to space constraints.

\section{Convergence Analysis}
This section analyzes the convergence behavior of the proposed coherence-aware distributed learning framework under mismatched coherence intervals, partial model updates via product superposition, and imperfect CSI.

\subsection{Preliminaries}
We aim to minimize the global loss function over $K$ participating devices, as defined in Eq.~\eqref{eq:global-loss}. Let $\btheta^* \triangleq \argmin_{\btheta \in \mathbb{R}^d} F(\btheta)$, and $F^* \triangleq F(\btheta^*)$. 
Further, define $\bA_k$ as a diagonal binary masking matrix for device~$k$, where $(\bA_k)_{ii} = 1$ if parameter~$i$ is received by device~$k$, and $0$ otherwise. Then, $\bA_k = \mathbf{I}, \forall k \in \mathcal{K}_S$.

The portion of the global model received by device~$k$ at round~$t$ is a noisy version of $\mathbf{A}_k \bar{\btheta}_k^{(t)}$, where $\bar{\btheta}_k^{(t)}$ denotes the imperfect estimate of the global model $\btheta^{(t)}$ at device~$k$ (see the description preceding Eq.~\eqref{eq:local-updates} for details). Let $\mathbf{e}_{k,S}^{(t)}$ and $\mathbf{e}_{k,D}^{(t)}$ denote the total noise in the decoded parameter vectors at static and dynamic devices, respectively. The total noise at dynamic devices is higher, as $\mathbf{e}_{k,D}^{(t)}$ includes both receiver AWGN and residual channel estimation error (see Eq.~\eqref{eq:estimation-error}), whereas $\mathbf{e}_{k,S}^{(t)}$ accounts only for receiver AWGN.

The initial local model for any static device~$k \in \mathcal{K}_S$ is set to $\btheta_{k,1}^{(t)} = \bar{\btheta}_k^{(t)} = \btheta^{(t)} + \be_{k,S}^{(t)}$. For any dynamic device~$k \in \mathcal{K}_D$, the initial model is constructed based on the partially received global model and a specific filling strategy to handle missing parameters. We consider two such strategies in this work:
\begin{itemize}[leftmargin=5mm]
\item \textbf{Zero-Filling (ZF):} The device sets unreceived parameters to zero, resulting in a projection of the received signal as
\begin{align}
    \btheta_{k,1}^{(t)} = \bA_k \btheta^{(t)} + \bA_k \be_{k,D}^{(t)}.
    \label{eq:ZF-model}
\end{align}

\item \textbf{Previous Local Model Filling (PLMF):} The device fills in missing parameters using its own final local model from the previous round, $\btheta_{k,\tau}^{(t-1)}$. Therefore, 
\begin{align}
         \btheta_{k,1}^{(t)} = \bA_k \btheta^{(t)} + \bA_k \be_{k,D}^{(t)} + (\bI - \bA_k) \btheta_{k,\tau}^{(t-1)}.  
 \label{eq:PreviousUpdates-model}
\end{align}
\end{itemize}

Each scheduled device $k \in [K]$ then performs $\tau$ steps of SGD on its local dataset, as given in Eq.~\eqref{eq:local-updates}.
After $\tau$ local steps, $\Delta \btheta_k^{(t)} = \btheta_{k,\tau}^{(t)} - \btheta_{k,1}^{(t)}$ is sent back to the PS from device~$k$ via a perfect uplink channel. The PS then aggregates these models to form the new global model for the next round using Eq.~\eqref{eq:model-update}.

\begin{remark}
 While our framework supports multiple strategies for handling unreceived model parameters—such as zero-filling and PLMF—we present the detailed convergence analysis only for the PLMF strategy, which is more complex than zero-filling. The analysis for the alternative strategy follows a similar structure, but is omitted for brevity due to space limitations. Nonetheless, experimental results in Section~\ref{sec:numerical-results} compare both strategies to validate their effectiveness within the proposed framework.
\end{remark}

\subsection{Assumptions}
Our analysis relies on the following standard assumptions~\cite{Cao2022PowerControl,Yan2024OTAFedAvg,Wang2022IRSFL,Zheng2023Semifed,Sun2022DynamicScheduling,Asaad2024JointAntenna}.

\noindent{\em Assumption 1:} Each local loss function $F_k, k \in [K]$, is $L$-smooth; that is, $\forall \boldsymbol{\phi}, \boldsymbol{\psi} \in \mathbb{R}^{d}$
    \begin{align*}
        F_k(\boldsymbol{\phi}) - F_k(\boldsymbol{\psi}) \le \langle \nabla F_k(\boldsymbol{\psi}), \boldsymbol{\phi} - \boldsymbol{\psi} \rangle + \frac{L}{2} \|\boldsymbol{\phi} - \boldsymbol{\psi}\|^2.
    \end{align*}
    Thus, the global loss function $F$ is also $L$-smooth.

\noindent{\em Assumption 2:} The variance of the stochastic gradients is bounded. For any device $k \in [K]$ and model $\btheta$,
    \begin{align*}
        \mathbb{E}\big[\|\nabla F_k(\btheta ; \bbeta) - \nabla F_k(\btheta)\|^2\big] \le \gamma^2.
    \end{align*}

\noindent{\em Assumption 3:} The downlink noise terms $\be_{k,S}^{(t)}$ (for static devices) and $\be_{k,D}^{(t)}$ (for dynamic devices) are zero-mean, and their variances are bounded:
    \begin{align*}
    \mathbb{E}\big[\|\be_k^{(t)}\|^2\big] \le
    \begin{cases}
     \sigma_S^2, & \text{if } k \in \mathcal{K}_S \\
     \sigma_D^2, & \text{if } k \in \mathcal{K}_D 
     \end{cases}.
    \end{align*}
    Note that $\sigma_{S}^2 < \sigma_{D}^2$, since static devices do not experience channel estimation errors (see Section~\ref{sec:downlink-signaling} for details).

\noindent{\em Assumption 4:} Local gradients may differ from the global gradient due to heterogeneous parameter distribution. For any $\btheta$, we have
    \begin{align*}
        \frac{1}{K} \sum_{k=1}^{K} \|\nabla F_k(\btheta) - \nabla F(\btheta)\|^2 \le \omega^2.
    \end{align*}

\noindent{\em Assumption 5:} For each device $k \in [K]$, the stochastic gradient is an unbiased estimator of the true local gradient. Therefore, $\mathbb{E}[\nabla F_k(\btheta ; \bbeta)] = \nabla F_k(\btheta)$.

\begin{theorem}
\label{theo:ZF-NonConvexFunctions}
Under Assumptions 1–5, for a non-convex L-smooth global loss function, if the learning rate is chosen such that $\eta_\ell \le \frac{1}{2L\tau}$, the product superposition FL scheme using PLMF over $T$ rounds of training satisfies
\begin{align*}
\frac{1}{T}\sum_{t=0}^{T-1}\mathbb{E}&\left[\|\nabla F(\btheta^{(t)})\|^{2}\right] \le \frac{4(F(\btheta^{(0)})-F^{*})}{T\eta_{g}}\\
+ &\frac{4L^2\tau\eta_g}{T} \sum_{t=0}^{T-1} \mathbb{E}\left[\|\btheta^{(t)}-\btheta^{(t-1)}\|^{2}\right] + Z,
\end{align*}
where $\eta_g = \eta_\ell\tau$ is the effective global learning rate (assuming $\eta_\ell = \eta_{k,i}, \forall k \in [K], \forall i \in [\tau]$), and the irreducible error floor $Z$ is defined as
\begin{align*}
Z = 8L\eta_g\tau(\gamma^2+\omega^2) + 4L\sigma_D^2.
\end{align*}
\end{theorem}
\proof \, See Appendix~\ref{app:proof-ZF-nonconvex}.

From Theorem~\ref{theo:ZF-NonConvexFunctions}, we can see how the convergence speed depends on the filling strategy for the missing parameters, especially in early rounds (see Eq.~\eqref{eq:PreviousUpdates-model}). For large $T$, the bound reduces to $Z$, where we see the impact of the dominant noise at the devices ($\sigma_D^2$). $\sigma_D^2$ originates from the dynamic devices in our scheme, and is dependent on our superposition pilots. This increases the overall noise in the system, but achieves another source of gain that improve the convergence behavior under coherence disparity.
   
\section{Numerical Experiments}
\label{sec:numerical-results}

\subsection{Simulation Setup}
Unless stated otherwise, we set $\rho = 10$ dB and use the power allocation calculated in Eqs.~\eqref{eq:rho_d_opt} and \eqref{eq:rho_p_opt}. The total noise used to compute the downlink SNR at static devices consists solely of receiver AWGN with variance $\sigma_w^2$ (see Section~\ref{sec:system-model}). For dynamic devices, the total noise includes both $\sigma_w^2$ and the channel estimation error introduced by product superposition, as given in Eq.~\eqref{eq:estimation-error}. The total pilot overhead during downlink communication is denoted by $\lambda \in [0,1]$, defined as the ratio of slots used for pilot transmission (either ordinary or superposed) to the total number of downlink communication slots. The value of $\lambda$ depends on the coherence times of dynamic links and the level of disparity among them. We denote the total number of communication rounds by $T$.

We conduct experiments using the MNIST~\cite{Lecun1998mnist} and CIFAR-10~\cite{krizhevsky2009learning} datasets.\footnote{We focus on relatively simple ML tasks here as a proof-of-concept of our innovations in addressing downlink impairments in distributed learning.} For training on MNIST, we use the default convolutional neural network (CNN) architecture with convolutional and fully connected layers. For CIFAR-10, we employ the ResNet-18 architecture. Each device performs local training using SGD for $\tau = 5$ local epochs with a batch size of 16. Both i.i.d. and non-i.i.d. data distributions are considered across the devices, as explained below.

\begin{figure}
    \centering
    \includegraphics[width=0.95\linewidth]{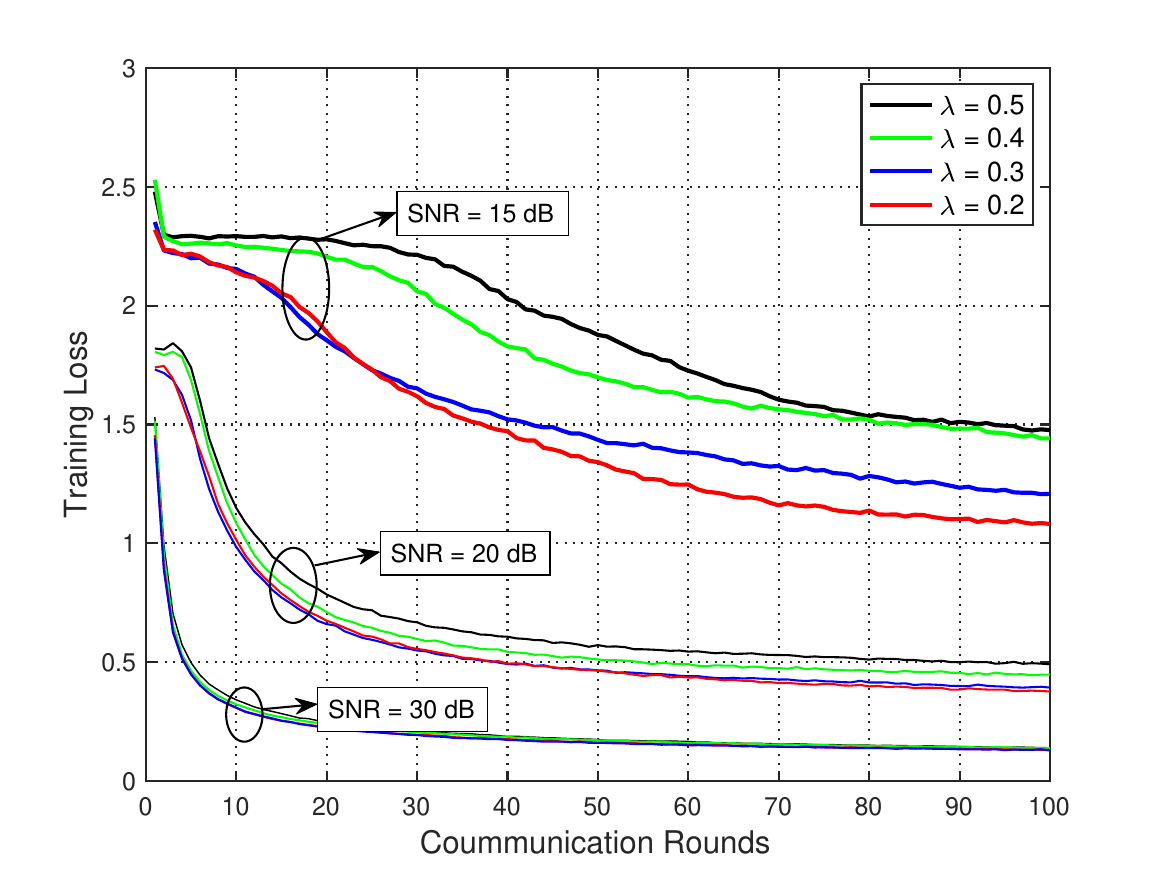}
    \caption{Training loss versus communication rounds on for the proposed product superposition-based FL scheme (MNIST). \vspace{-0.15in}}
    \label{fig:train-loss}
\end{figure}

\subsection{Results and Discussion}
Fig.~\ref{fig:train-loss} demonstrates the training loss of FL using the proposed product superposition scheme under different SNR and $\lambda$ values for the MNIST dataset with i.i.d. distribution. Here, we set $M = 20$, $K = 50$, and $|\mathcal{K}_S| = |\mathcal{K}_D| = 25$. The results confirm that product superposition is a valid approach for FL under varying coherence disparities, which result in different $\lambda$ values. The learning performance improves significantly at higher SNRs and with lower pilot overhead, which is due to the reduced noise from dynamic devices under these conditions.

\begin{figure*}[ht]
    \centering

    \begin{subfigure}{0.32\textwidth}
        \centering
        \includegraphics[width=\linewidth]{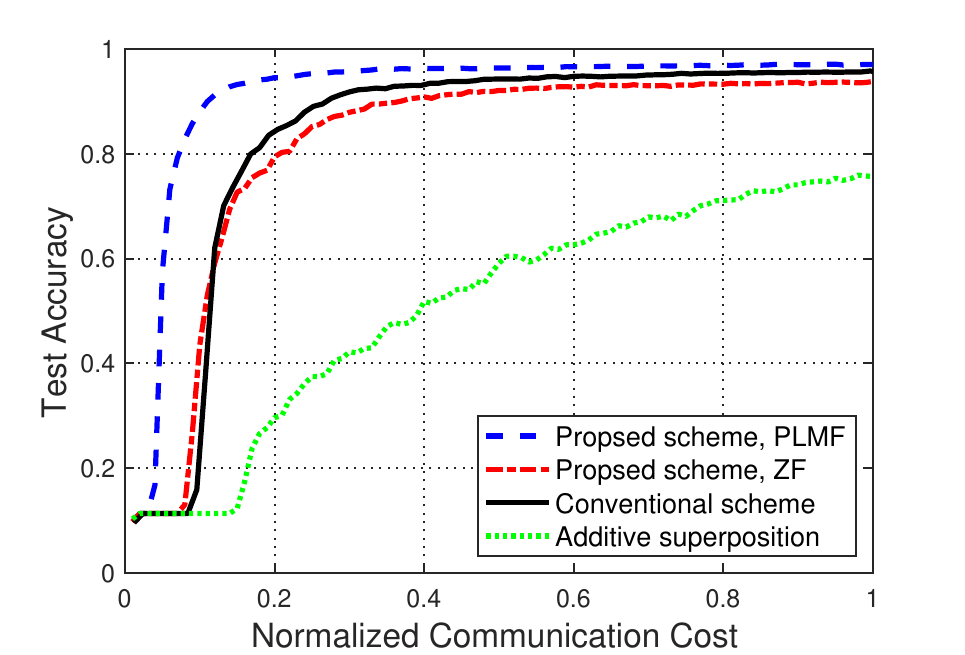}
        \caption{$\lambda = 0.2$}
        \label{fig:subfig1}
    \end{subfigure}
    \hfill
    \begin{subfigure}{0.32\textwidth}
        \centering
        \includegraphics[width=\linewidth]{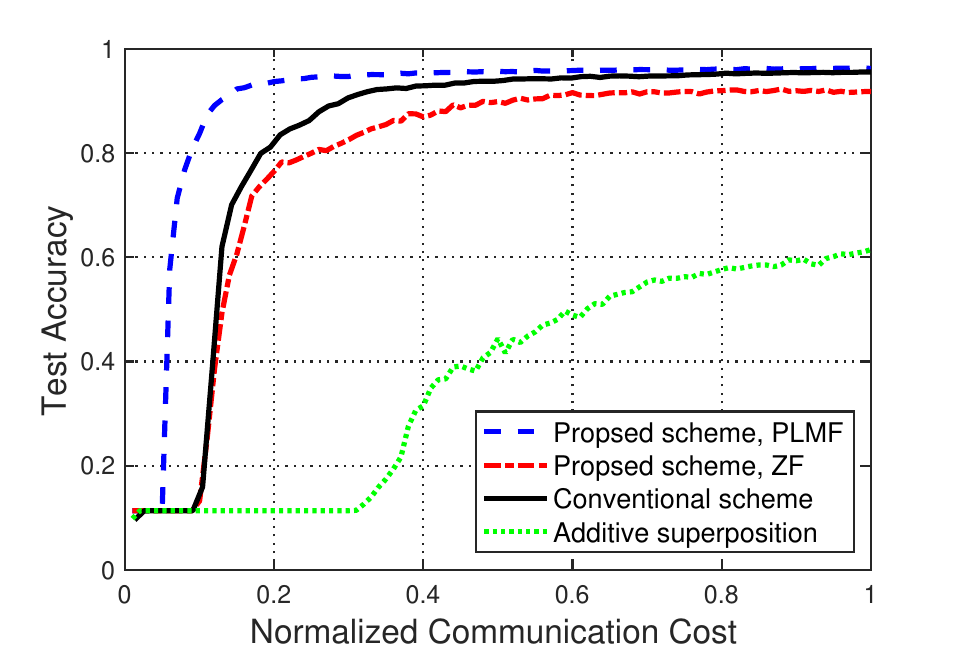}
        \caption{$\lambda = 0.3$}
        \label{fig:subfig2}
    \end{subfigure}
    \hfill
    \begin{subfigure}{0.32\textwidth}
        \centering
        \includegraphics[width=\linewidth]{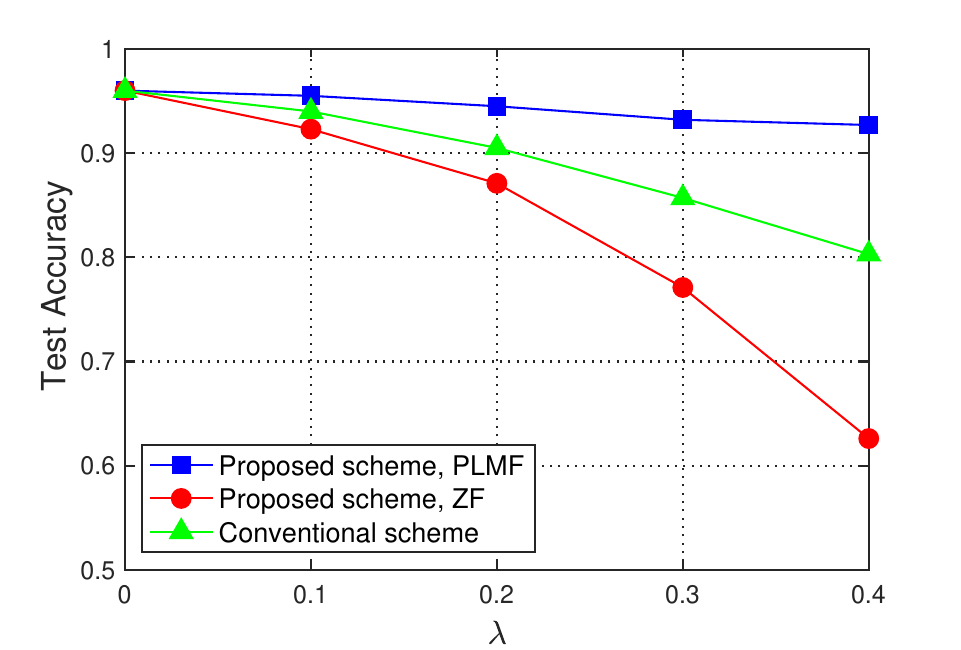}
        \caption{$T = 20$}
        \label{fig:subfig3}
    \end{subfigure}

    \caption{Test accuracy comparison between the proposed scheme and conventional baselines, for the MNIST dataset. (a) and (b) show test accuracy versus normalized communication cost at $\lambda = 0.2$ and $\lambda = 0.3$, respectively. (c) presents test accuracy as a function of $\lambda$ with a fixed training duration of $T = 20$. \vspace{-0.15in}}
    \label{fig:snr20test-accuracy}
\end{figure*}

Fig.~\ref{fig:snr20test-accuracy}(a) and Fig.~\ref{fig:snr20test-accuracy}(b) show the test accuracy versus the normalized communication cost, defined as the ratio of total slots required for downlink communication (including pilot and parameter transmissions) to the total slots required for parameter transmission alone, at $\lambda = 0.2$ and $\lambda = 0.3$, respectively. The plot compares the proposed product superposition scheme with benchmark methods under the MNIST dataset with i.i.d. distribution. Here, we set $M = 20$, $\rm{SNR} = 20$ dB, $K = 50$, $|\mathcal{K}_S| = |\mathcal{K}_D| = 25$, and $T = 100$. The proposed scheme with PLMF significantly outperforms conventional FL, which uses conventional signaling (orthogonal pilot and parameter transmission) for model delivery, yielding substantial gains in communication efficiency. This improvement is due to the efficient resource management enabled by product superposition—particularly the optimized pilot placement and reuse—which reduces communication overhead while maintaining high test accuracy. The use of zero-filling to handle missing parameters at dynamic devices degrades the test accuracy, as it increases bias in the learning process.

Another benchmark is the additive superposition scheme, where pilot and parameter signals are added under the coherence disparity. While this method allows for pilot reuse for parameter transmission, it performs poorly because the superimposed pilot acts as interference to parameters, introducing additional noise in the decoded parameters and degrading learning performance. In contrast, our product superposition approach addresses this limitation by integrating the pilot signal into a {\em virtual channel} estimated at dynamic devices (see Eq.~\eqref{eq:MMSE-equivalent}) without any interference parameters. This is one of the key advantages that make the proposed method suitable for FL under coherence disparity.

Overall, the proposed scheme with PLMF outperforms all baselines. For instance, at 95\% test accuracy, it achieves a normalized communication cost reduction of approximately 0.3 compared to conventional FL.

Fig.~\ref{fig:snr20test-accuracy}(c) shows the test accuracy versus the pilot overhead under the same setting, with a fixed training duration of $T = 20$. When all devices are static, i.e., $\lambda = 0$, all schemes perform similarly. However, as $\lambda$ increases—reflecting greater coherence disparity—the performance of both the conventional scheme and the product superposition method with zero-filling degrades significantly. In contrast, the proposed scheme with PLMF remains robust, demonstrating its effectiveness under heterogeneous coherence conditions. At $\lambda = 0.4$, the product superposition with PLMF achieves approximately a 0.12 improvement in test accuracy over the baseline.

\begin{figure}[t]
    \centering
    \begin{subfigure}{\linewidth}
        \centering
        \includegraphics[width=0.9\linewidth]{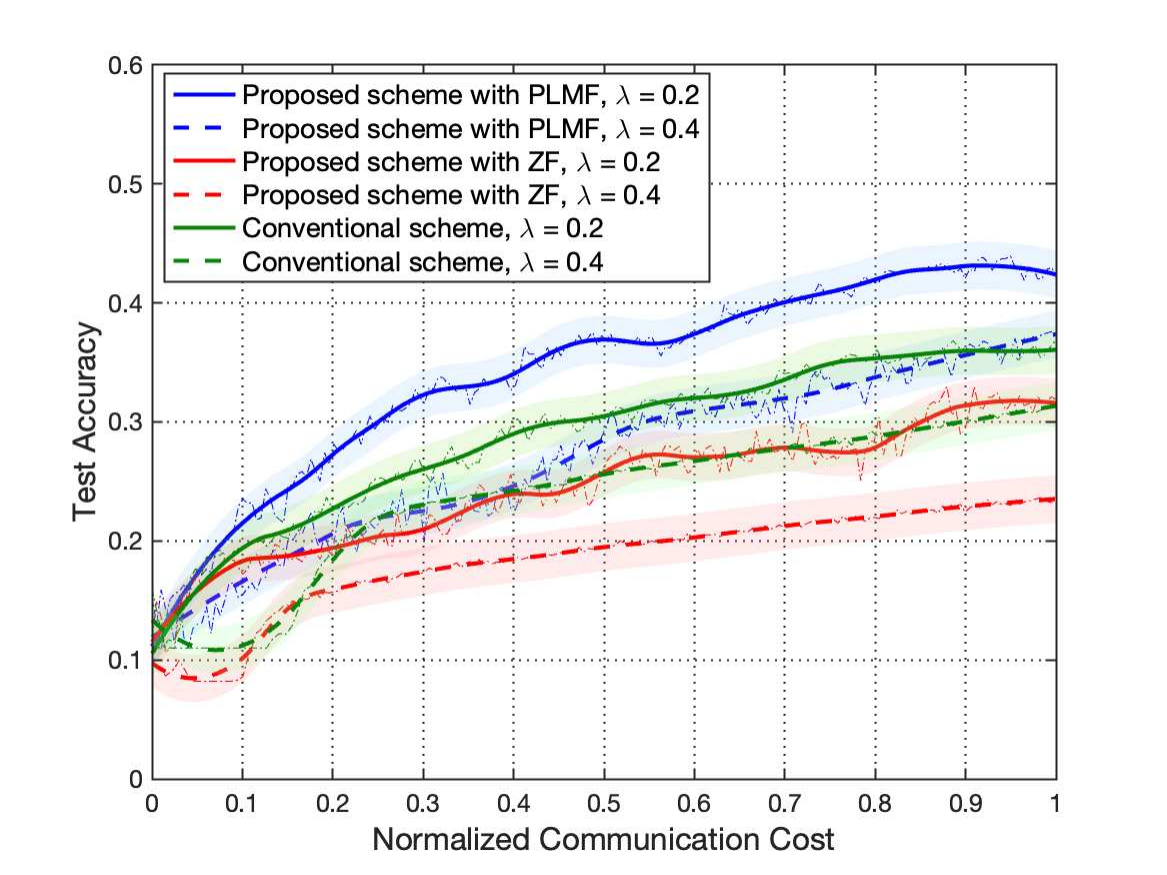}
        \caption{SNR = 10 dB}
        \label{fig:cifar-snr10}
    \end{subfigure}
    \begin{subfigure}{\linewidth}
        \centering
        \includegraphics[width=0.9\linewidth]{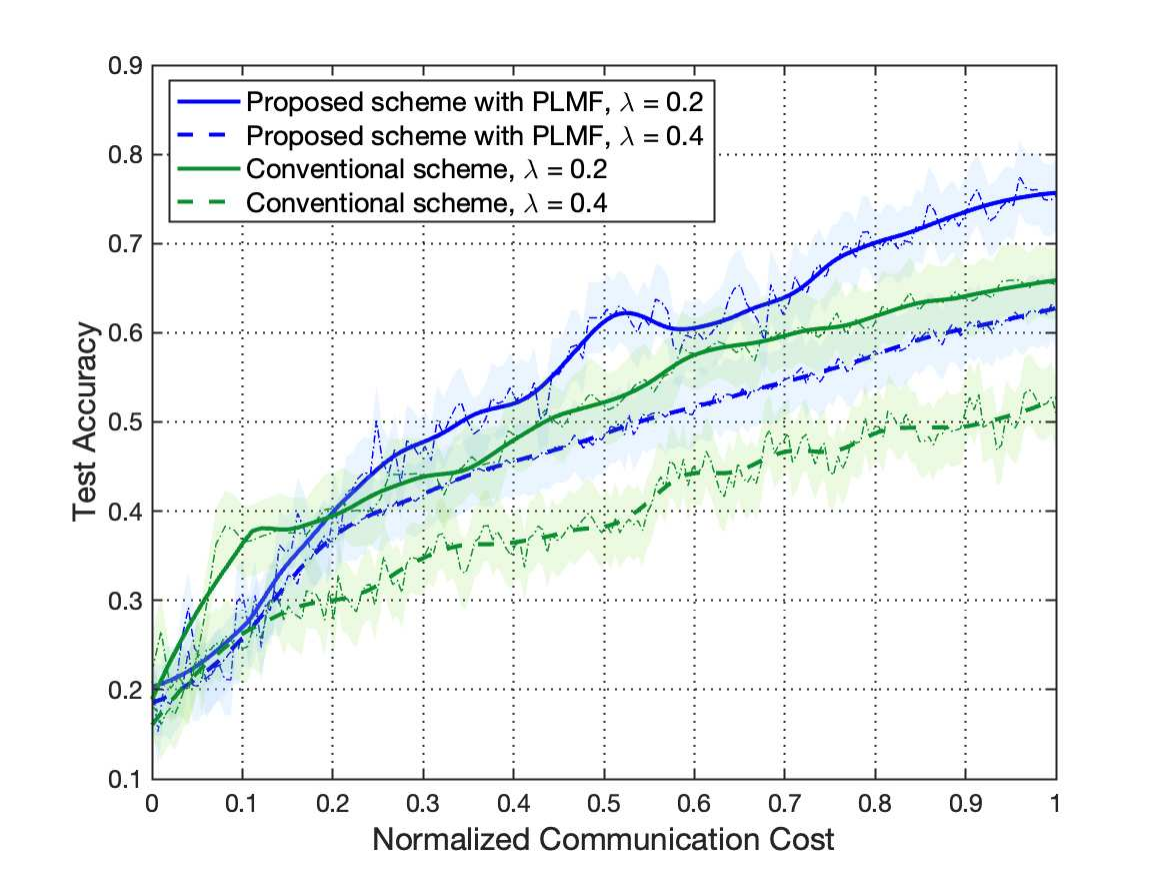}
        \caption{SNR = 30 dB}
        \label{fig:cifar-resnet-snr30}
    \end{subfigure}    
    \caption{Test accuracy versus normalized communication cost on the CIFAR-10 dataset for the proposed product superposition-based FL, and the conventional FL with ordinary pilots: (a) SNR = 10 dB, and (b) SNR = 30 dB. Shaded regions indicate the standard deviation of the test accuracy. \vspace{-0.15in}}
    \label{fig:test-accuracy-snr30}
\end{figure}


Fig.~\ref{fig:test-accuracy-snr30} compares the test accuracy of FL under the proposed signaling scheme and conventional signaling on the CIFAR-10 dataset across different communication rounds and pilot overheads. Here, we set $M = 30$, $\mathrm{SNR} = \{10, 30\}$ dB, $K = 40$ with $|\mathcal{K}_S| = 0.6K$ and $|\mathcal{K}_D| = 0.4K$, $\lambda = \{0.2, 0.4\}$, $T=100$, assuming a non-i.i.d. data distribution. The proposed product superposition scheme with the PLMF strategy consistently outperforms the other approaches, achieving significant gains in communication efficiency under coherence disparity. In particular, when $\mathrm{SNR} = 30$ dB, at a test accuracy of 66\%, it achieves approximately a 0.28 reduction in normalized communication cost compared to conventional FL.

\section{Conclusion}
This paper proposed coherence-aware FL, addressing a key limitation in the assumption of uniform channel conditions across devices for downlink model delivery. In practice, the performance of FL critically depends on the availability of accurate and timely CSI, which becomes particularly challenging in networks with heterogeneous coherence times. Coherence disparity leads to unequal channel training requirements and inefficient resource utilization, which can significantly degrade FL performance. To tackle this, we proposed a methodology based on product superposition that jointly handles downlink pilot signaling and model broadcasting. This design allows dynamic devices to estimate virtual channels while enabling static devices to receive full global updates through pilot reuse, significantly improving communication efficiency without requiring additional spectrum or signaling overhead. Simulation results confirmed that the proposed method outperforms conventional and additive-superposition FL baselines.

We focused on the challenges introduced by downlink impairments, including partial model reception and imperfect channel state information. To isolate these effects, we adopted a simplified FDD uplink with perfect CSI, enabling focused analysis of downlink-induced degradation. While this work assumes a basic aggregation strategy, the proposed framework remains compatible with more sophisticated uplink models, which are left for future exploration.

\appendices

\section{Pilot-Parameter Power Allocation}
\label{app:proof-PowerAllocation}
\subsection{Achievable Rate by Static Device over the Pilot Slots}
A static device $k'$ has perfect knowledge of $\bh_{k'}$, and the unitary pilot matrix, $\bX_p \in \mathbb{C}^{M \times M}$. During the pilot transmission phase (the first $M$ time slots), the signal received by user $k'$ is
\[
\by_{k',p} = \sqrt{\rho_p}\, \bh_{k'}^H \bX_{p}^\theta \bX_p + \bw_{k',p},
\]
where $\bX_{p}^\theta \in \mathbb{C}^{M \times M}$ is the parameter matrix, and $\bw_{k',p}$ is the AWGN.

To decode the parameter matrix $\bX_{p}^\theta$, the device right-multiplies the received signal by the conjugate transpose of the known pilot matrix, $\bX_p^H$. Since $\bX_p$ is unitary ($\bX_p\bX_p^H = \bI_M$), this operation effectively removes the pilot modulation
\begin{align*}
    \by'_{k',p} &= \by_{k',p} \,\bX_p^H \\
&=\sqrt{\rho_p} \,\bh_{k'}^H \bX_{p}^\theta (\bX_p \bX_p^H) + \bw_{k',p} \bX_p^H \\
&= \sqrt{\rho_p} \,\bh_{k'}^H \bX_{p}^\theta + \bw'_{k',p}.
\end{align*}
The resulting noise term, $\bw'_{k',p} \triangleq \bw_{k',p} \bX_p^H$, has the same statistical properties as the original noise. The equation above describes a standard $M \times 1$ MISO channel. The capacity, assuming i.i.d. inputs from the $M$ antennas, is given by $\mathbb{E}[\log_2(1 + \frac{\rho_p}{M\sigma_w^2} \bh_{k'}^H \bh_{k'})]$. Averaging this rate over the entire block of $T_K$ symbols gives the final expression for $R_{k'}$.

\subsection{Dynamic Device Rate and Effective SNR}
During the pilot phase, the dynamic device~$k$ estimates its {\em virtual channel}, which is defined as the product of the physical channel and the parameter matrix: $\bf_{k} = \bh_k^H \bX_\theta$. Its MMSE estimate, denoted $\overline{\bf}_{k}$, is given in Eq.~\eqref{eq:MMSE-equivalent}, and the associated estimation error is calculated in Eq.~\eqref{eq:estimation-error}. A key property of MMSE estimation is that the estimate $\overline{\bf}_{k}$ and the error $\tilde{\bf}_{k}$ are uncorrelated~\cite{Hassibi2003HowMuch}. Let $\alpha^2 \triangleq \frac{M\rho_p}{\sigma_w^2+M\rho_p}$. Then 
\begin{align*}
\mathbb{E}\big[||\tilde{\bf}_{k}||^2\big] &= \mathbb{E}\big[||\bf_{k}||^2\big] - \mathbb{E}\big[||\overline{\bf}_{k}||^2\big] = M(1-\alpha^2)
\end{align*}

During the data transmission phase, the received signal at a specific time slot is $\by_{k,d} = \sqrt{\rho_d}\, \bf_{k} \bX_d + \bw_d$. We substitute $\bf_{k} = \overline{\bf}_{k} + \tilde{\bf}_{k}$ to separate the signal from the effective noise
\[
\by_{k,d} = \underbrace{\sqrt{\rho_d} \,\overline{\bf}_{k} \bX_d^\theta}_{\text{signal}} + \underbrace{\sqrt{\rho_d}\, \tilde{\bf}_{k} \bX_d^\theta + \bw_d}_{\text{effective noise}}.
\]
The instantaneous SNR is the ratio of the signal power (conditioned on the estimate $\overline{\bf}_{k}$) to the variance of the effective noise. The signal power is $\rho_d ||\overline{\bf}_{k}||^2$. The noise variance is $\sigma_w^2 + \mathbb{E}[||\sqrt{\rho_d}\, \tilde{\bf}_{k} \bx_d||^2] = \sigma_w^2 + \sqrt{\rho_d}\,\mathbb{E}[||\tilde{\bf{k}}||^2] = \sigma_w^2 + M\rho_d(1-\alpha^2)$~\cite{Hassibi2003HowMuch}. The instantaneous SNR is therefore
\begin{align*}
   \text{SNR}_k &= \frac{\rho_d ||\overline{\bf}_{k}||^2}{\sigma_w^2 + M\rho_d(1-\alpha^2)} \\
   &= \left(\frac{\rho_d(\sigma_w^2+M\rho_p)}{\sigma_w^2(\sigma_w^2+M\rho_p+M\rho_d)}\right) ||\overline{\bf}_{k}||^2. 
\end{align*}
The achievable rate for the dynamic device~$k$, $R_k$, is found by taking the expectation of $\log_2(1+\text{SNR}_k)$ over the distribution of the channel estimate $\overline{\bf}_{k}$. Therefore, the effective SNR is
\[
\gamma_{\text{eff},k} = \frac{\rho_d(\sigma_w^2+M\rho_p)}{\sigma_w^2(\sigma_w^2+M\rho_p+M\rho_d)}.
\]

\subsection{Optimal Power Allocation Derivation}

The objective is to maximize the dynamic device's rate by maximizing $\gamma_{\text{eff},k}$ subject to the total power constraint, which we assume is met with equality
\[
M \big( \rho_p + \rho_d(T_K - M) \big) = \rho \frac{s}{q} = \rho T_K.
\]
Maximizing $\gamma_{\text{eff},k}$ is equivalent to minimizing its reciprocal, $g(\rho_p, \rho_d) = \frac{\sigma_w^2}{\rho_d} + \frac{\sigma_w^2M}{\sigma_w^2+M\rho_p}$. From the power constraint, we express $\rho_p$ in terms of $\rho_d$. Let the constant $c = \frac{\rho T_K}{M}$. Then $\rho_p = c - \rho_d(T_K-M)$. Substitute this into $g(\rho_p, \rho_d)$
\[
g(\rho_d) = \frac{\sigma_w^2}{\rho_d} + \frac{\sigma_w^2M}{\sigma_w^2 + M(c - \rho_d(T_K-M))}.
\]
To find the minimum, we take the derivative with respect to $\rho_d$ and set it to zero. This results in
\[
\frac{\sigma_w^2}{\rho_d^2} = \frac{\sigma_w^2M^2(T_K-M)}{(\sigma_w^2+Mc-M(T_K-M)\rho_d)^2}.
\]
Taking the square root and rearranging to solve for $\rho_d$ yields the optimal allocation $\rho_d^*$
\begin{align*}
    \rho_d^* &= \frac{\sigma_w^2+Mc}{M\sqrt{T_K-M}(1 + \sqrt{T_K-M})} \\
    &= \frac{\sigma_w^2+\rho T_K}{M\sqrt{T_K-M}(1 + \sqrt{T_K-M})}.
\end{align*}
The optimal pilot power, $\rho_p^*$, is found by substituting $\rho_d^*$ back into the power constraint equation. This completes the proof.

\section{Proof of Theorem~\ref{theo:ZF-NonConvexFunctions}}
\label{app:proof-ZF-nonconvex}
From Assumption~1, we have the fundamental descent lemma for the global update $\Delta\btheta^{(t)} = \btheta^{(t+1)} - \btheta^{(t)}$ as
\begin{align} 
\mathbb{E}[F(\btheta^{(t+1)})]& \le  \mathbb{E}[F(\btheta^{(t)})] \nonumber \\
+& \mathbb{E}[\langle\nabla F(\btheta^{(t)}), \Delta\btheta^{(t)}\rangle]+ \frac{L}{2}\mathbb{E}[\|\Delta\btheta^{(t)}\|^{2}].
\label{eq:descent_lemma}
\end{align}
Our proof strategy is to bound the last two terms in the inequality above. Through extensive yet standard algebraic manipulations, based on Assumptions 1–5 and the PLMF update rule in Eq.~\eqref{eq:PreviousUpdates-model}, we derive the following bounds.

First, we bound the inner product term. Using the definition of $\Delta\btheta^{(t)}$ and taking the expectation with respect to the stochastic noise, we have
\begin{align*}
    &\mathbb{E}[\langle\nabla F(\btheta^{(t)}), \Delta\btheta^{(t)}\rangle] \\&= -\eta_\ell \sum_{i=0}^{\tau-1} \mathbb{E}\left[\left\langle\nabla F(\btheta^{(t)}), \mathbb{E}_k[\nabla F_k(\btheta_{k,i}^{(t)})]\right\rangle\right] \\
    &= -\eta_g \mathbb{E}[\|\nabla F(\btheta^{(t)})\|^2] \\
    &- \eta_\ell \sum_{i=0}^{\tau-1} \mathbb{E}\left[\left\langle\nabla F(\btheta^{(t)}), \mathbb{E}_k[\nabla F_k(\btheta_{k,i}^{(t)}) - \nabla F(\btheta^{(t)})]\right\rangle\right].
\end{align*}
Applying $2\langle a,b \rangle \leq \|a\|^2 + \|b\|^2$ to the second term on the right-hand side yields
\begin{align*}
    \mathbb{E}[\langle\nabla F(\btheta^{(t)}),& \Delta\btheta^{(t)}\rangle] 
    \le -\frac{\eta_g}{2} \mathbb{E}[\|\nabla F(\btheta^{(t)})\|^2] \\
    +& \frac{\eta_g}{2\tau} \sum_{i=1}^{\tau}\mathbb{E}[\|\mathbb{E}_k[\nabla F_k(\btheta_{k,i}^{(t)}) - \nabla F(\btheta^{(t)})]\|^2].
\end{align*}
We add and subtract $\nabla F_k(\btheta^{(t)})$ inside the norm, and then apply the inequality $\|a+b\|^2 \leq 2\|a\|^2+2\|b\|^2$ to obtain
\begin{align*}
    &\mathbb{E}[\|\mathbb{E}_k[\nabla F_k(\btheta_{k,i}^{(t)}) - \nabla F(\btheta^{(t)})]\|^2] \\
    &\le 2\mathbb{E}[\|\mathbb{E}_k[\nabla F_k(\btheta^{(t)}) - \nabla F(\btheta^{(t)})]\|^2] \\
    &+ 2\mathbb{E}[\|\mathbb{E}_k[\nabla F_k(\btheta_{k,i}^{(t)}) - \nabla F_k(\btheta^{(t)})]\|^2].
\end{align*}
The first part is bounded by Assumption~4 as
\begin{align*}
    2\mathbb{E}[\|\mathbb{E}_k[\nabla F_k(\btheta^{(t)}) - \nabla F(\btheta^{(t)})]\|^2] \le 2\omega^2, 
\end{align*}
and the second term is bounded by Assumption~1 as
\begin{align*}
    2\mathbb{E}[\|\mathbb{E}_k[\nabla F_k(\btheta_{k,i}^{(t)}) - \nabla F_k(\btheta^{(t)})]\|^2] \leq 2L^2 \mathbb{E}[\|\btheta_{k,i}^{(t)} - \btheta^{(t)}\|^2].
\end{align*}
Next, we must bound the local model drift term, $\mathbb{E}[\|\btheta_{k,i}^{(t)} - \btheta^{(t)}\|^2]$. This drift depends on the initial model error at the start of the round and the accumulation of local updates. A standard derivation shows that the average drift is bounded as
\begin{align*}
    \frac{1}{\tau}\sum_{i=1}^{\tau}\mathbb{E}[\|\btheta_{k,i}^{(t)}-\btheta^{(t)}\|^2] &\le 2\mathbb{E}[\|\btheta_{k,1}^{(t)} - \btheta^{(t)}\|^2] \\
    &+ 2\eta_\ell^2\tau^2(\gamma^2+\omega^2).
\end{align*}
The initial model error itself, $\mathbb{E}[\|\btheta_{k,1}^{(t)} - \btheta^{(t)}\|^2]$, is bounded by analyzing the PLMF update rule. This error contains the effects of using a stale model and the downlink noise, leading to the bound
\begin{align*}
    \mathbb{E}[\|\btheta_{k,1}^{(t)} - \btheta^{(t)}\|^2] &\le 2\mathbb{E}[\|\btheta^{(t)} - \btheta^{(t-1)}\|^2] \\
    &+ 2\eta_g^2(\gamma^2+\omega^2) + \sigma_D^2.
\end{align*}
By substituting these nested bounds back into the inequality for the inner product, we establish its final bound as
\begin{align*}
\mathbb{E}[\langle\nabla F(\btheta^{(t)}), \Delta\btheta^{(t)}\rangle] &\le -\frac{\eta_g}{2} \mathbb{E}[\|\nabla F(\btheta^{(t)})\|^2] \\&+ 2L^2\tau\eta_g^2 \mathbb{E}[\|\btheta^{(t)}-\btheta^{(t-1)}\|^2] \\
&+ 4L^2\eta_g^3\tau(\gamma^2+\omega^2) + L^2\eta_g\tau\sigma_D^2.
\end{align*}
Second, we bound the squared update norm term, $\frac{L}{2}\mathbb{E}[\|\Delta\btheta^{(t)}\|^{2}]$. Following a similar process of decomposing the variance, $\mathbb{E}[\|X\|^2] = \|\mathbb{E}[X]\|^2 + {\rm Var}(X)$, and bounding the local drift terms using Assumptions~3 and 4, we find
\begin{align*}
\frac{L}{2}\mathbb{E}[\|\Delta\btheta^{(t)}&\|^2] \le L\eta_g^2(\gamma^2+\omega^2) \\
&+ L^3\eta_g^2\Big(\frac{1}{\tau}\sum_i \mathbb{E}[\|\btheta_{k,i}^{(t)}-\btheta^{(t)}\|^2]\Big).
\end{align*}
Substituting these bounds into Eq.~\eqref{eq:descent_lemma} and simplifying under the learning rate condition $\eta_\ell \le \frac{1}{2L\tau}$, we arrive at the following inequality for a single round $t$
\begin{align}
\frac{\eta_g}{4}\mathbb{E}[\|\nabla F(\btheta^{(t)})\|^2] 
&\le \mathbb{E}[F(\btheta^{(t)})] - \mathbb{E}[F(\btheta^{(t+1)})]  \nonumber\\
 + &L^2\tau\eta_g^2 \mathbb{E}[\|\btheta^{(t)}-\btheta^{(t-1)}\|^2] \nonumber\\
 +& (2L\eta_g^2\tau)(\gamma^2+\omega^2) + (L\eta_g)\sigma_D^2.
\label{eq:single_round}
\end{align}
We now sum Eq.~\eqref{eq:single_round} over all communication rounds from $t=1$ to $t=T$. We have
\begin{align*}
\sum_{t=1}^{T} \frac{\eta_g}{4}\mathbb{E}[\|\nabla F(\btheta^{(t)})\|^2] &\le \sum_{t=1}^{T} (\mathbb{E}[F(\btheta^{(t)})] - \mathbb{E}[F(\btheta^{(t+1)})]) \\
&+ \sum_{t=0}^{T-1} \Big[L^2\tau\eta_g^2 \mathbb{E}[\|\btheta^{(t)}-\btheta^{(t-1)}\|^2] \nonumber\\
&+ 2L\eta_g^2\tau(\gamma^2+\omega^2) + L\eta_g\sigma_D^2\Big].
\end{align*}
The first term on the right-hand side is a telescoping sum:
\begin{align*}
\sum_{t=1}^{T} (\mathbb{E}[F(\btheta^{(t)})] - \mathbb{E}[F(\btheta^{(t+1)})]) &= \mathbb{E}[F(\btheta^{(0)})] - \mathbb{E}[F(\btheta^{(T)})] \\
& \le F(\btheta^{(0)}) - F^*
\end{align*}
Substituting this and dividing by $T(\eta_g/4)$, we get
\begin{align*}
&\frac{1}{T}\sum_{t=0}^{T-1}\mathbb{E}[\|\nabla F(\btheta^{(t)})\|^2] \le \frac{4(F(\btheta^{(0)}) - F^*)}{T\eta_g} \\
&+ \frac{4}{T\eta_g}\sum_{t=0}^{T-1} \left( L^2\tau\eta_g^2 \mathbb{E}[\|\btheta^{(t)}-\btheta^{(t-1)}\|^2] + \dots \right) \\
&= \frac{4(F(\btheta^{(0)})-F^{*})}{T\eta_{g}} + \frac{4L^2\tau \eta_g}{T} \sum_{t=0}^{T-1} \mathbb{E}[\|\btheta^{(t)}-\btheta^{(t-1)}\|^{2}] \\
& \quad + 8L\eta_g\tau(\gamma^2+\omega^2) + 4L\sigma_D^2
\end{align*}
This completes the proof.

\clearpage

\balance
\bibliographystyle{IEEEtran}
\bibliography{IEEEabrv,ref}

\end{document}